\documentclass[twocolumn,showpacs,preprintnumbers,amsmath,
                amssymb,prb,eqsecnum]{revtex4}

\usepackage{graphicx}
\usepackage{dcolumn}
\usepackage{bm}
\usepackage{longtable}%

\begin{document}

\preprint{E2/ver.4}

\title{A theory of the electric quadrupole contribution
to resonant x-ray scattering:\\
Application to multipole ordering phases in Ce$_{1-x}$La$_x$B$_6$}

\author{Tatsuya Nagao}
\affiliation{%
Faculty of Engineering, Gunma University, Kiryu, Gunma 376-8515, Japan}

\author{Jun-ichi Igarashi}%
\affiliation{%
Faculty of Science, Ibaraki University, Mito, Ibaraki 310-8512, Japan}

\date{\today}

\begin{abstract}
We study the electric quadrupole ($E$2) contribution
to resonant x-ray scattering (RXS).
Under the assumption that the rotational invariance is
preserved in the Hamiltonian describing the intermediate state of 
scattering, we derive a useful expression for the RXS amplitude. 
One of the advantages the derived expression possesses is the full 
information of the energy dependence, lacking in all the 
previous studies using the fast collision approximation.
The expression is also helpful to classify the spectra
into multipole order parameters which are brought about.
The expression is suitable to investigate the RXS
spectra in the localized $f$ electron systems.
We demonstrate the usefulness of the formula by calculating 
the RXS spectra at the Ce $L_{2,3}$ edges in Ce$_{1-x}$La$_{x}$B$_{6}$
on the basis of the formula. We obtain the spectra as a function of 
energy in agreement with the experiment of Ce$_{0.7}$La$_{0.3}$B$_{6}$. 
Analyzing the azimuthal angle dependence, we find 
the sixfold symmetry in the $\sigma$-$\sigma'$ channel and the threefold one 
in the $\sigma$-$\pi'$ channel not only in the antiferrooctupole (AFO) 
ordering phase but also in the antiferroquadrupole (AFQ) ordering phase,
which behavior depends strongly on the domain distribution.
The sixfold symmetry in the AFQ phase arises from the simultaneously induced
hexadecapole order.
Although the AFO order is plausible for phase IV in Ce$_{1-x}$La$_{x}$B$_{6}$,
the possibility of the AFQ order may not be ruled out on the basis
of azimuthal angle dependence alone.
\end{abstract}

\maketitle

\section{\label{sect.1}Introduction}

Resonant x-ray scattering (RXS) has recently attracted much attention,
since strong x-ray intensities have become available from the synchrotron 
radiation. It is described by a second-order process that a core electron
is excited into unoccupied states by absorbing incident x-rays and 
that electron is recombined with the core hole by emitting x-rays.
The RXS has been recognized as a useful probe to investigate spatially 
varying multipole orderings, which the conventional neutron scattering is 
usually difficult to detect.

For probing the spacial variation of order parameters, 
x-ray wavelengths need to be order of the variation period.
In transition metals, the $K$ edges in the dipole ($E$1) transition 
are just fitting for this purpose. 
Actually, by using the $K$ edge, the possibility of the orbital ordering 
has already been explored in transition-metal compounds.
\cite{Murakami98,Murakami98.2}
The RXS intensities are observed at superlattice Bragg spots,
which are interpreted as originating from the modulation in the $4p$ band,
since the process involves the excitation of a $1s$ electron to unoccupied
$4p$ states.

Because the ordering pattern is usually controlled by electrons in the
$3d$ band, the mechanism which causes the modulation
is not necessarily trivial. 
Actually, for most of transition-metal compounds,
both experimental studies and theoretical studies based
on electronic structure calculations have revealed that
the RXS intensities are brought about by the hybridization
between the $4p$ band and the $2p$ band of the neighboring anions 
rather than the direct Coulomb interaction between the electron
in the $4p$ band and electrons in the $3d$ band.\cite{Ohsumi03,Igarashi05}
This result is reasonable because of the extended nature of the $4p$ state.

On rare earth metal compounds such as CeB$_{6}$, DyB$_{2}$C$_{2}$, 
the $L_{2,3}$ edges in the $E$1 transition are used because of 
the requirement for x-ray wavelength.
\cite{Nakao01,Yakhou01,Hirota00,Tanaka99} 
The RXS spectra in the $E$1 transition from the antiferroquadrupole
(AFQ) phase of CeB$_{6}$ were studied 
both experimentally\cite{Nakao01} and theoretically.\cite{Nagao01,Igarashi02} 
Although the experiments and the theory give sufficiently consistent results,
the relation to the multipole orderings which $4f$ electrons 
mainly involve is rather indirect, 
since the resonance is caused by the excitation of a $2p$ electron 
to $5d$ states. This shortcoming may be overcome by using the quadrupole
($E$2) transition 
at the $L_{2,3}$ edges, where a $2p$ electron is promoted to partially 
filled $4f$ states. Using the $E$2 transition has another merit that
octupole and hexadecapole orderings are directly detectable.
This contrasts with the $E$1 transition, where 
only dipole and quadrupole orderings are detectable. 
Of course, intensities in the $E$2 transition are usually
much smaller than those in the $E$1 transition.

In this paper, we derive a general formula of the RXS amplitudes in the 
localized electron picture, in which the electronic structure at each atom 
is assumed to be well described by an atomic wavefunction under 
the crystal electric field
(CEF). Historically, the research in such a direction was started by using 
the framework borrowed from resonant $\gamma$-ray scattering.\cite{Trammel62}
Starting from the works by Blume and Gibbs\cite{Blume88} and by Hannon
\textit{et al}.\cite{Hannon88},
the form of the RXS amplitude had been investigated
in several works.\cite{Luo93,Carra94,Lovesey96,Hill96}
The RXS amplitude can be summarized into an elegant form
by using vector spherical harmonics.  Unfortunately,
it has little practical usage because it is difficult
to deduce meaningful information when there is no restriction
on the intermediate state of the scattering process.
A widely-adopted approximation for practical use is the so-called 
"fast collision (FC) approximation". This replaces the intermediate state 
energy in the energy denominator of the RXS amplitude 
by an averaged value, allowing the denominator out of the summation.
\cite{Hannon88,Luo93,Carra94,Lovesey96,Hill96}
Thereby, the multiplet splitting of the intermediate state is neglected,
leading to an assumed form (usually a Lorentzian form) for the energy profile. 

However, recent experiments show deviation from the Lorentzian form 
in several materials.\cite{Paixao02,Wilkins04}
We improve the situation
by taking the energy dependence of the intermediate state 
under the assumption that the intermediate Hamiltonian 
describing the scattering process preserves spherical symmetry.
This assumption is justified when the CEF energy and the intersite interaction 
are much smaller than the multiplet energy in the intermediate state 
as is expected in many localized electron materials.
We have already reported the formula for the $E$1 transition, 
having successfully applied to the analysis of the $E$1 RXS spectra in
URu$_{2}$Si$_{2}$ and NpO$_{2}$.\cite{Nagao05,Nagao05.2,Nagao06}
This paper is an extension of those works to the $E$2 transition.
The obtained formula makes it possible to analyze the energy profiles 
of the spectra in contrast with the FC approximation.
In addition, the formula is suitable to discuss the relation of
the RXS spectra to multipole order parameters,
\cite{Isaacs90,Paixao02,Bernhoeft03,Wilkins04,Mannix05,Wilkins06}
because it is expressed by means of the expectation values
of the multipole order parameters. 

We demonstrate the usefulness of the formula by calculating the RXS spectra
in multipole ordering phases of Ce$_{1-x}$La$_{x}$B$_{6}$.
First, we investigate the $E$2 RXS spectra at the Ce $L_{2,3}$ edges 
from the AFQ ordering phase (phase II)
in the non-diluted material CeB$_{6}$. Analysis utilizing our formula
reveals that the $E$2 RXS spectra in phase II consist of a mixture
of the quadrupole and hexadecapole energy profiles.
The calculated intensities suggest the possibility that
he $E$2 signal at the Ce $L_{2,3}$ edges
can be detectable in this material.

For the intermediate doping range $x \approx 0.3 \sim 0.5$,
Ce$_{1-x}$La$_{x}$B$_{6}$ falls into a new phase (phase IV) whose
primary order parameter is not well established yet.\cite{Tayama97}
Recently, the $E$2 RXS signals at the Ce $L_2$ edge have been detected 
for an $x=0.3$ sample.\cite{Mannix05}
From the azimuthal angle ($\psi$) dependence of the peak intensity,
it was claimed that the antiferrooctupole (AFO) ordering phase is
the most probable candidate because the symmetry of the
$\psi$-dependence, sixfold and threefold in the $\sigma$-$\sigma'$ 
and $\sigma$-$\pi'$ channels respectively, is deduced
from the theory in good agreement with the experiment.\cite{Mannix05,Kusunose05}
However, the relative intensity between two channels depends strongly on
the domain distribution, and deviates about factor two from the experimental 
one if the contribution from four domains are added with equal weight.
The origin of this discrepancy is still unanswered.
It will be pointed out that the RXS peak intensity from the AFQ phase 
concomitant with the induced hexadecapole contribution also gives rise to
the same symmetry of the $\psi$-dependence as obtained from the AFO phase.
Thus, although the AFO order is plausible in many respects,
it seems difficult to rule out the AFQ order on the basis of the azimuthal
angle dependence alone. 
In addition, we calculate the energy dependence 
of the RXS spectra at the Ce $L_{2,3}$ edges.
Assuming both the AFO and AFQ orders, we obtain the spectral shapes
at the $L_2$ edge, which agree with the experimental one 
for Ce$_{0.7}$La$_{0.3}$B$_{6}$.\cite{Mannix05} 
On the other hand, the spectral shapes  at the $L_3$ edge are found 
slightly different between two phases, with
intensities the same order of magnitude of the reported one at the $L_2$ edge.

The present paper is organized as follows.
A general expression for RXS amplitudes is obtained in Sec. \ref{sect.2}.
Analysis of RXS spectra in Ce$_{1-x}$La$_{x}$B$_{6}$ 
are presented in Sec. \ref{sect.3}. 
Section \ref{sect.4} is devoted to concluding remarks.
In Appendix, we show several expressions required to obtain the
RXS amplitude formula.

\section{\label{sect.2}Theoretical Framework of RXS}

\subsection{a second-order optical process}

\begin{figure}
\includegraphics[width=8.0cm]{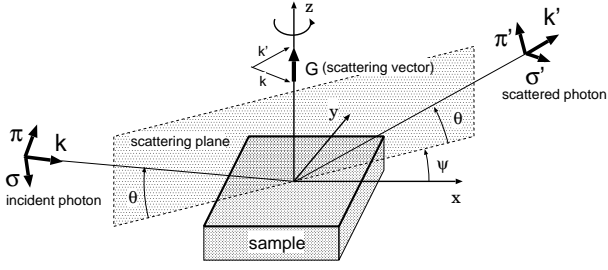}
\caption{\label{fig.geom}
Geometry of the RXS experiment. Photon with polarization 
$\sigma$ or $\pi$ is scattered into the state of polarization
$\sigma'$ or $\pi'$ at the Bragg angle $\theta$.
The azimuthal angle $\psi$ describes the rotation of the sample around 
the scattering vector ${\textbf G}$.
The $(11\overline{2})$ surface is in the scattering plane at $\psi=0$.
}
\end{figure}

The RXS is described by a second-order optical process,
where a core electron is excited to unoccupied states by absorbing x-rays
and that electron is recombined with the core hole by emitting x-rays.
Since the wavefunction of core electron is well localized,
the RXS amplitude may be given by a sum of contributions from individual ions. 
Using a geometrical arrangement shown in Fig. \ref{fig.geom},
we express the RXS amplitude 
$f(\mbox{\boldmath{$\epsilon$}},\mbox{\boldmath{$\epsilon$}}',
{\textbf k}, {\textbf k}',\omega)$ for the incident x-ray with momentum
${\textbf k}$, polarization $\mbox{\boldmath{$\epsilon$}}$, and
the scattered x-ray with momentum ${\textbf k}'$, polarization
$\mbox{\boldmath{$\epsilon$}}'$ as
\begin{eqnarray}
 f(\mbox{\boldmath{$\epsilon$}},\mbox{\boldmath{$\epsilon$}}'
  ,{\textbf k}, {\textbf k}',\omega) &=&
\sum_{n=1} f^{(n)}(\mbox{\boldmath{$\epsilon$}},\mbox{\boldmath{$\epsilon$}}'
   ,{\textbf k}, {\textbf k}',\omega), \\
 f^{(n)}(\mbox{\boldmath{$\epsilon$}},\mbox{\boldmath{$\epsilon$}}'
   ,{\textbf k}, {\textbf k}',\omega) &\propto&
   \frac{1}{\sqrt{N}}\sum_{j} {\textrm e}^{-i{\textbf G}\cdot
{\textbf r}_j} 
\nonumber \\
&\times& 
 M_j^{(n)}(\mbox{\boldmath{$\epsilon$}},\mbox{\boldmath{$\epsilon$}}',
    {\textbf k}, {\textbf k}',\omega), 
\label{eq.rxs.intensity}
\end{eqnarray}
with
\begin{eqnarray}
 M_j^{(1)}(\mbox{\boldmath{$\epsilon$}},\mbox{\boldmath{$\epsilon$}}',\omega) 
 &=&\sum_{\mu,\mu'}\epsilon'_{\mu} \epsilon_{\mu'} 
\nonumber \\
&\times&
  \sum_{\Lambda} 
  \frac{\langle\psi_0|x_{\mu, j} |\Lambda\rangle
  \langle \Lambda|x_{\mu', j}|\psi_0\rangle}
       {\hbar\omega-(E_{\Lambda}-E_0)+i\Gamma},
\label{eq.amplitude.dipole} \\
 M_j^{(2)}(\mbox{\boldmath{$\epsilon$}},\mbox{\boldmath{$\epsilon$}}',
 {\textbf k},{\textbf k}',\omega) 
 &=& \frac{k^2}{9}\sum_{\mu,\mu'} 
  q_{\mu}(\hat{\bf k}',\mbox{\boldmath{$\epsilon$}}')
  q_{\mu'}(\hat{\bf k},\mbox{\boldmath{$\epsilon$}})
\nonumber \\
&\times&
  \sum_{\Lambda} 
  \frac{\langle\psi_0|\tilde{z}_{\mu, j} |\Lambda\rangle
  \langle \Lambda|\tilde{z}_{\mu', j}|\psi_0\rangle}
       {\hbar\omega-(E_{\Lambda}-E_0)+i\Gamma}, 
\label{eq.amplitude.quadrupole} 
\end{eqnarray}
where ${\textbf G}$ ($={\textbf k}'-{\textbf k}$) is the scattering vector,
and $N$ is the number of sites $j$'s.
The $| \psi_0 \rangle$ represents the ground state with energy $E_0$, 
while $|\Lambda\rangle$ represents the intermediate state with energy
$E_{\Lambda}$. The $\Gamma$ describes the life-time broadening width 
of the core hole.
Equation (\ref{eq.amplitude.dipole}) describes the $E$1 transition,
where the dipole operators $x_{\mu, j}$'s are defined as
$x_{1, j}=x_{j}$, $x_{2, j}=y_{j}$, and $x_{3, j}=z_{j}$ 
in the coordinate frame fixed to the crystal axes with the origin located 
at the center of site $j$.
Equation (\ref{eq.amplitude.quadrupole}) describes the $E$2 transition,
where the quadrupole operators are defined by
$\tilde{z}_{1,j}=\frac{\sqrt{3}}{2}(x_j^2-y_j^2)$,
$\tilde{z}_{2,j}=\frac{1}{2}(3z_j^2-r_j^2)$,
$\tilde{z}_{3,j}=\sqrt{3}y_jz_j$, $\tilde{z}_{4,j}=\sqrt{3}z_jx_j$
and $\tilde{z}_{5,j}=\sqrt{3}x_jy_j$.
Factors $q_{\mu}(\hat{\bf k},\mbox{\boldmath{$\epsilon$}})$
and $q_{\mu}(\hat{\bf k}',\mbox{\boldmath{$\epsilon$}}')$
with $\hat{\textbf k} = {\textbf k}/|{\textbf k}|$
and $\hat{\textbf k}' = {\textbf k}'/|{\textbf k}'|$
are defined as a second-rank tensor,
\begin{equation}
 q_{\mu}({\textbf A},{\textbf B})
\equiv \left\{ \begin{array}{ll}
 \frac{\sqrt{3}}{2} \left( A_x B_x - A_y B_y \right)  & \textrm{for} \ \mu=1, \\
\frac{1}{2} \left( 3 A_z B_z - {\textbf A} \cdot {\textbf B} \right) 
                                                      & \textrm{for} \ \mu=2, \\
 \frac{\sqrt{3}}{2} \left( A_y B_z + A_z B_y \right) & \textrm{for} \ \mu=3, \\
 \frac{\sqrt{3}}{2} \left( A_z B_x + A_x B_z \right) & \textrm{for} \ \mu=4, \\
 \frac{\sqrt{3}}{2} \left( A_x B_y + A_y B_x \right) & \textrm{for} \ \mu=5. \\
\end{array} \right. 
\end{equation}
Note that the quadrupole operators $\tilde{z}_{\mu,j}$ are expressed as
$\tilde{z}_{\mu,j}= q_{\mu}({\textbf r}_j, {\textbf r}_j)$.

\subsection{Energy profiles}

In localized electron systems, the ground state and the intermediate 
state are described in terms of the eigenfunctions of the angular momentum 
operator, $|J,m\rangle$, at each site.  At the ground state,
the CEF and the intersite interaction usually lift the degeneracy 
with respect to $m$. We write the ground state at site $j$ as
\begin{equation}
  |\psi_0\rangle_j = \sum_m c_j(m)|J,m\rangle.
\label{eq.ground}
\end{equation}
In the intermediate state, however, the CEF and the intersite interaction 
may be neglected in a good approximation, 
since their magnitudes of energy are much 
smaller than those of the intra-atomic Coulomb interaction 
and the spin-orbit interaction (SOI) 
which give rise to the multiplet structure.
Thus the Hamiltonian describing the intermediate state is approximated
as preserving the spherical symmetry. 
In such a circumstance, the intermediate states are characterized by
the total angular momentum at the core-hole site, that is,
$|\Lambda \rangle=|J',M,i \rangle$ with the magnitude $J'$ 
and the magnetic quantum number $M$. The corresponding energy is denoted by
$E_{J',i}$, where we introduce the index $i$ in order to distinguish 
multiplets having the same $J'$ value but having different energy.

In the following, we discuss only on the $E$2 transition
(Eq.~(\ref{eq.amplitude.quadrupole})),
because the $E$1 transition has been fully analyzed 
in our previous paper.\cite{Nagao05.2}
First, we rewrite Eq.~(\ref{eq.amplitude.quadrupole}) as
\begin{eqnarray}
 M_j^{(2)}(\mbox{\boldmath{$\epsilon$}},\mbox{\boldmath{$\epsilon$}}',
 {\textbf k},{\textbf k}',\omega) 
 &=&\frac{k^2}{9}\sum_{\mu,\mu'} 
q_{\mu}(\hat{\textbf k}',\mbox{\boldmath{$\epsilon$}}')
q_{\mu'}(\hat{\textbf k},\mbox{\boldmath{$\epsilon$}})
M_{\mu\mu'}^{(2)}(\omega,j), \nonumber \\
\label{eq.amplitude.E2.1.5} \\
M_{\mu\mu'}^{(2)}(\omega,j) &\equiv&
  \sum_{J',M,i} E_{i}(\omega,J')
  \langle\psi_0|\tilde{z}_{\mu, j} |J',M,i\rangle \nonumber \\
&\times&
  \langle J',M,i|\tilde{z}_{\mu', j}|\psi_0\rangle,
\label{eq.amplitude.E2.2}
\end{eqnarray}
with
\begin{equation}
E_i(\omega,J') = \frac{1}
       {\hbar\omega-(E_{J',i}-E_0)+i\Gamma}.
\end{equation}
Then, inserting Eq.~(\ref{eq.ground}) for the ground state into 
Eq.~(\ref{eq.amplitude.E2.2}), we obtain
\begin{equation}
 {M}_{\mu \mu'}^{(2)}(j,\omega)
 =  \sum_{m,m'} c^{\ast}_j(m) c_j(m') 
   {M}_{\mu \mu'}^{(2) m,m'}(\omega), \label{eq.amplitude.3}
\end{equation}
with
\begin{eqnarray}
 {M}_{\mu \mu'}^{(2) m,m'}(\omega)
  &=& \sum_{J'} \sum_{i=1}^{N_{J'}} E_i(\omega,J')
\nonumber \\
& & \hspace*{-2.20cm}\times
    \sum_{M=-J'}^{J'} 
  \langle J,m|\tilde{z}_{\mu}|J',M,i \rangle
  \langle J',M,i|\tilde{z}_{\mu'}| J,m'\rangle.
    \label{eq.amplitude.E2.4} 
\end{eqnarray}
We have suppressed the index $j$ specifying the core-hole site.
The number of the multiplets having the value $J'$ is denoted by $N_{J'}$.
The selection rule for the $E$2 transition confines the range of the summation
over $J'$ to $J'=J, J \pm 1, J \pm 2$. 

Now we analyze the matrix element of the type
$\langle J, m|\tilde z_\mu|J', M \rangle$ by utilizing the
Wigner-Eckart (WE) theorem for a tensor operator,\cite{Messiah62}
\begin{eqnarray}
 \langle J,m|s_\mu|J'M\rangle &=& (-1)^{J'+m-2} \sqrt{2J+1} \nonumber \\
&\times&
\left( \begin{array}{ccc}
J' & 2 & J \\
M & \mu & -m 
\end{array} \right) (J||V_2||J'),
\end{eqnarray}
with $s_{\pm 2} = (\tilde{z}_1 \pm {\textrm i} \tilde{z}_5)/\sqrt{2}$,
$s_{\pm 1} = \mp (\tilde{z}_4 \pm {\textrm i} \tilde{z}_3)/\sqrt{2}$
and $s_0=\tilde{z}_2$.
The symbol $(J||V_2||J')$ denotes the
reduced matrix element of the set of irreducible tensor operator of the 
second rank. 
Because of the nature of the quadrupole operators,
a condition $|m -m '| \leq 4$ has to be satisfied for non-vanishing
${M}_{\mu\mu'}^{(2) m,m '}(\omega)$.
After a straightforward but tedious calculation with the help of
the WE theorem, we obtain non-zero ${M}^{(2) m,m '}(\omega)$'s.
Then, we perform the summation over $m$ and $m'$ in Eq. (\ref{eq.amplitude.3}).
The result is summarized by introducing the
expectation values of the components of the
multipole operators as follows:
\begin{equation}
M_{\mu,\mu'}^{(2)}(j,\omega) =
\sum_{\nu=0}^{4} \alpha_{E2}^{(\nu)}(\omega) \sum_{\lambda=1}^{2 \nu+1}
[ M_{\lambda}^{(\nu)} ]_{\mu,\mu'}
\langle \psi_0 | z_{\lambda}^{(\nu)} | \psi_0 \rangle,
\label{eq.amp.new}
\end{equation}
where the $\lambda$ th component of rank $\nu$ tensor $z_{\lambda}^{(\nu)}$
in real basis ($1 \leq \lambda \leq 2 \nu+1$) is defined in
Table \ref{table.4}. 
The $z_{\lambda}^{(\nu)}$ is constructed from the irreducible spherical 
tensor $T_{\lambda}^{(n)}$ through the unitary transformation $U^{(\nu)}$.
The definitions of $T_{n}^{(\nu)}$ and $U^{(\nu)}$ as well as
the energy profile $\alpha_{E2}^{(\nu)}(\omega)$
are given in Appendix \ref{app.A}.
The matrix element of $M_{\lambda}^{(\nu)}$ is expressed as
\begin{eqnarray}
 [ M_{\lambda}^{(\nu)}]_{\mu, \mu'}
&=&\frac{(-)^{\nu}}{(2||T_{\nu}||2)} \sqrt{\frac{2\nu+1}{5}}
\sum_{\ell=-2}^{2} \sum_{\ell'=-2}^{2}
U_{\mu \ell}^{(2)} \nonumber \\
&\times&
\sum_{n=-\nu}^{\nu} U_{\lambda n}^{(\nu)} 
([T_{n}^{(\nu)}]_{\ell \ell'})^{\ast}
[U^{(2) \dagger}]_{\ell' \mu'}
\nonumber \\
&=&(-)^{\nu} \sqrt{2\nu+1} \sum_{\ell=-2}^{2} \sum_{\ell'=-2}^{2} 
(-)^{\ell} U_{\mu \ell}^{(2)} \nonumber \\
&\times&
\sum_{n=-\nu}^{\nu} U_{\lambda n}^{(\nu)} 
\left( \begin{array}{ccc}
 2 & \nu & 2 \\
 \ell' & n & -\ell \\
 \end{array} \right)
[U^{(2) \dagger}]_{\ell' \mu'},
\nonumber \\
\label{eq.m.matrix}
\end{eqnarray}
with
\begin{equation}
(2||T_{\nu}||2) = \frac{1}{2^{\nu}} \sqrt{\frac{(5+\nu)!}{5(4-\nu)!}}.
\end{equation}

\squeezetable
\begin{table}
\caption{
\label{table.4} 
Definition of the operator equivalence of the multipole order components.
The overline denotes the symmetrization, for instance,
$\overline{X^2 Y}=X^2 Y + XYX + YX^2$.
}
\begin{ruledtabular}
\begin{tabular}{l}
$z_{1}^{(1)} =J_x$ \\
$z_{2}^{(1)} =J_y$ \\
$z_{3}^{(1)} =J_z$ \\
\hline
$z_{1}^{(2)} = O_{x^2-y^2} = \frac{\sqrt{3}}{2}[J_x^2-J_y^2]$ \\
$z_{2}^{(2)} = O_{3z^2-r^2} = \frac{1}{2}[3 J_z^2-J(J+1)]$ \\
$z_{3}^{(2)} = O_{yz} = \frac{\sqrt{3}}{2}[ J_y J_z + J_z J_y]$ \\
$z_{4}^{(2)} = O_{zx} = \frac{\sqrt{3}}{2}[ J_z J_x + J_x J_z]$ \\
$z_{5}^{(2)} = O_{xy} = \frac{\sqrt{3}}{2}[ J_x J_y + J_y J_x]$ \\
\hline
$z_{1}^{(3)} =
   T_{xyz} = \frac{\sqrt{15}}{6} \overline{J_{x}J_{y}J_{z}}$ \\
$z_{2}^{(3)} =
   T_{x}^{\alpha} = \frac{1}{2}[2J_x^3- \overline{J_x(J_y^2+J_z^2)}]$ \\
$z_{3}^{(3)} =
   T_{y}^{\alpha} = \frac{1}{2}[2J_y^3- \overline{J_y(J_z^2+J_x^2)}]$ \\
$z_{4}^{(3)} =
   T_{z}^{\alpha} = \frac{1}{2}[2J_z^3- \overline{J_z(J_x^2+J_y^2)}]$  \\
$z_{5}^{(3)} =
   T_{x}^{\beta} = \frac{\sqrt{15}}{6} \overline{J_{x}(J_{y}^2-J_{z}^2)}$ \\
$z_{6}^{(3)} =
   T_{y}^{\beta} = \frac{\sqrt{15}}{6} \overline{J_{y}(J_{z}^2-J_{x}^2)}$ \\
$z_{7}^{(3)} =
   T_{z}^{\beta} = \frac{\sqrt{15}}{6} \overline{J_{z}(J_{x}^2-J_{y}^2)}$ \\
\hline
$z_{1}^{(4)} =
   H_{4}^{0} = \frac{5}{4} \sqrt{\frac{7}{3}} 
      \left[ J_x^4+J_y^4+J_z^4 
    -\frac{3}{5} J(J+1) 
     \left\{ J(J+1)-\frac{1}{3} \right\} \right]$  \\
$z_{2}^{(4)}=
   H_{4}^{2} =-\frac{\sqrt{5}}{4} 
      \left[ \frac{7}{6} \overline{(J_x^2-J_y^2)J_z^2} 
   -\left\{J(J+1)-\frac{5}{6} \right\} (J_x^2-J_y^2) \right]$ \\
$z_{3}^{(4)}=
   H_{4}^{4} = \frac{\sqrt{5}}{48} 
      \left[ 35J_z^4-30J(J+1)J_z^2 
  +3J(J+1) [J(J+1)-2] \right.$ \\
 \hspace*{1.5cm}$\left.+25 J_z^2 -7 [J_x^4+J_y^4-\overline{J_x^2J_y^2} ] 
    \right]$ \\
$z_{4}^{(4)}=
   H_{x}^{\alpha} = \frac{\sqrt{35}}{8}
     [\overline{J_y^3 J_z} - \overline{J_y J_z^3} ]$ \\
$z_{5}^{(4)}=
   H_{y}^{\alpha} = \frac{\sqrt{35}}{8}
     [\overline{J_z^3 J_x} - \overline{J_z J_x^3} ]$ \\
$z_{6}^{(4)}=
   H_{z}^{\alpha} = \frac{\sqrt{35}}{8}
     [\overline{J_x^3 J_y} - \overline{J_x J_y^3} ]$ \\
$z_{7}^{(4)}=
   H_{x}^{\beta} = \frac{\sqrt{5}}{8} 
[2\overline{J_x^2 J_y J_z} 
   -( \overline{J_y^3 J_z} + \overline{J_y J_z^3} ) ]$ \\
$z_{8}^{(4)}=
   H_{y}^{\beta} = \frac{\sqrt{5}}{8} 
[2\overline{J_y^2 J_z J_x} 
   -( \overline{J_z^3 J_x} + \overline{J_z J_x^3} ) ]$ \\
$z_{9}^{(4)}=
   H_{z}^{\beta} = \frac{\sqrt{5}}{8} 
[2\overline{J_z^2 J_x J_y} 
   -( \overline{J_z^3 J_y} + \overline{J_x J_y^3} ) ]$ \\
\end{tabular}
\end{ruledtabular}
\end{table}

Finally, inserting Eq.~(\ref{eq.amp.new}) into
Eq. (\ref{eq.amplitude.E2.1.5}) and using Eq.~(\ref{eq.m.matrix}), 
we obtain the final expression,
\begin{eqnarray}
& &  M_j^{(2)}(\mbox{\boldmath{$\epsilon$}},\mbox{\boldmath{$\epsilon$}}',
{\textbf k},{\textbf k}',\omega) 
\nonumber \\
 &=&\frac{k^2}{9}
\sum_{\nu=0}^{4} \alpha_{E2}^{(\nu)}(\omega) 
\sum_{\lambda=1}^{2\nu+1}
P_{\lambda}^{(\nu)} 
 (\mbox{\boldmath{$\epsilon$}},\mbox{\boldmath{$\epsilon$}}',
\hat{\textbf k},\hat{\textbf k}') 
\langle \psi_0 | z_{\lambda}^{(\nu)} | \psi_0 \rangle,
\nonumber \\
\label{eq.amplitude.E2.final}
\end{eqnarray}
where $P_{\lambda}^{(\nu)}$'s are the geometrical factors
defined as
\begin{eqnarray}
& & P_{\lambda}^{(\nu)}
 (\mbox{\boldmath{$\epsilon$}},\mbox{\boldmath{$\epsilon$}}',
\hat{\textbf k},\hat{\textbf k}') 
\nonumber \\
&=& \sqrt{2 \nu+1} 
\sum_{n=-\nu}^{\nu} (-)^{n} U_{\lambda n}^{(\nu)} \sum_{m=-\nu}^{\nu}
\left( \begin{array}{ccc}
  2 & 2 & \nu \\
  m & n-m & - n \\
 \end{array} \right) \nonumber \\
&\times& 
 q_{m}(\mbox{\boldmath{$\epsilon$}}',\hat{\textbf k}') 
 q_{n-m}(\mbox{\boldmath{$\epsilon$}},\hat{\textbf k}).
\end{eqnarray}

Those for $\nu=0,1$ and $2$ are expressed as relatively simple forms:
\begin{eqnarray}
 P_{1}^{(0)}(\mbox{\boldmath{$\epsilon$}},\mbox{\boldmath{$\epsilon$}}'
            ,{\textbf k},{\textbf k}')
&=& \frac{1}{\sqrt{5}} \left[ (\hat{\textbf k}' \cdot \hat{\textbf k}) 
(\mbox{\boldmath{$\epsilon$}}' \cdot \mbox{\boldmath{$\epsilon$}})
+ (\hat{\textbf k}' \cdot \mbox{\boldmath{$\epsilon$}})
(\mbox{\boldmath{$\epsilon$}}' \cdot \hat{\textbf k}) \right],
\nonumber \\
\\
P_{\mu}^{(1)}(\mbox{\boldmath{$\epsilon$}},\mbox{\boldmath{$\epsilon$}}'
             ,\hat{\textbf k},\hat{\textbf k}')
&   & \nonumber \\
& & \hspace{-1.70cm} =
 - {\textrm i} \frac{1}{\sqrt{10}} \left[ 
      (\mbox{\boldmath{$\epsilon$}}' \cdot \mbox{\boldmath{$\epsilon$}})
      (\hat{\textbf k}' \times \hat{\textbf k})_{\mu}
+ (\hat{\textbf k}' \cdot \hat{\textbf k})
  ( \mbox{\boldmath{$\epsilon$}}' \times \mbox{\boldmath{$\epsilon$}})_{\mu}
\right. \nonumber \\
& & \left. \hspace*{-1.50cm}
 + ( {\textrm k}' \cdot \mbox{\boldmath{$\epsilon$}})
  ( \mbox{\boldmath{$\epsilon$}}' \times \hat{\textbf k})_{\mu}
 +( \mbox{\boldmath{$\epsilon$}}' \cdot \hat{\textbf k})
  ( \hat{\textbf k}' \times \mbox{\boldmath{$\epsilon$}})_{\mu},
\right],
\\
P_{\mu}^{(2)}(\mbox{\boldmath{$\epsilon$}},\mbox{\boldmath{$\epsilon$}}'
             ,\hat{\textbf k},\hat{\textbf k}')
&=&  - \frac{3}{2} \frac{1}{\sqrt{14}} \left[
  (\mbox{\boldmath{$\epsilon$}}' \cdot \mbox{\boldmath{$\epsilon$}})
   q_{\mu}(\hat{\textbf k},\hat{\textbf k}') 
\right.
\nonumber \\
& & \left. \hspace*{-1.50cm}
+ (\hat{\textbf k}' \cdot \hat{\textbf k})  
   q_{\mu}(\mbox{\boldmath{$\epsilon$}},\mbox{\boldmath{$\epsilon$}}')
+ q_{\mu}(\hat{\textbf k}' \times \hat{\textbf k}, 
  \mbox{\boldmath{$\epsilon$}}' \times \mbox{\boldmath{$\epsilon$}})
\right].
\end{eqnarray}
For $\nu=1$, indices $\mu=1,2$ and $3$ serve as the Cartesian 
components $x,y$ and $z$, respectively.
The corresponding expression of $P_{\mu}^{(\nu)}$'s for $\nu=3,4$ 
have complicated forms, which are summarized in Appendix B.

An expression similar to Eq.~(\ref{eq.amplitude.E2.final}) 
has been derived by the FC approximation.
\cite{Hannon88,Luo93,Carra94,Hill96,Lovesey96}
However, this scheme has to put by hand the energy dependence.
The present theory gives an explicit expression of the energy dependence,
which is separated from the factor relating to the order parameter.
Thus, the choice of the CEF parameters in the ground state
does not affect the shape of energy profiles $\alpha_{E2}^{(\nu)}(\omega)$.

\section{\label{sect.3} Application to multipole ordering phases
in C\lowercase{e}$_{1-x}$L\lowercase{a}$_{x}$B$_{6}$}

In this section, we demonstrate the usefulness of 
Eq.(\ref{eq.amplitude.E2.final}) by analyzing the RXS spectra 
in the $E$2 transition at the Ce $L_{2,3}$ edges from 
Ce$_{1-x}$La$_{x}$B$_{6}$. 

\subsection{Phase II in CeB$_{6}$}

The parent material CeB$_{6}$ experiences two-step phase transitions.
It undergoes the first transition from paramagnetic (phase I)
to an AFQ state (phase II) at $T_{\textrm Q}=3.2$ K and
the second transition to an antiferromagnetic (AFM) state (phase III) 
at $T_{\textrm N}=2.4$ K under no external magnetic field.
The AFQ order is known to be a N\'{e}el-type with a propagating vector 
${\bf Q}_0=\left(\frac{1}{2} \frac{1}{2} \frac{1}{2}\right)$.

These phase transitions have been theoretically studied 
in a localized electron scheme, where each Ce ion is assumed 
to be trivalent in the $4f^1$ configuration. Its ground multiplet is a
$\Gamma_8$ quartet confined within the $J=\frac{5}{2}$ subspace 
under the cubic symmetry.
Using states $\{ |J_z=m \rangle \}$, the four bases 
$|\pm, \sigma \rangle$ ($\sigma =\uparrow, \downarrow$) may be expressed as
\begin{eqnarray}
 \left| +, \uparrow \right\rangle &=& 
 \sqrt{\frac{5}{6}} \left| +\frac{5}{2} \right\rangle 
+\sqrt{\frac{1}{6}} \left| -\frac{3}{2} \right\rangle, \\
 \left| -, \uparrow \right\rangle &=& \left| +\frac{1}{2} \right\rangle, 
\end{eqnarray}
and $|\pm, \downarrow \rangle$ by replacing $|m \rangle$
with $|-m \rangle$. 
The intersite interaction may lift the fourfold degeneracy,
leading to multipole orderings. Shiina \textit{et al}. have derived such
interaction from a microscopic model and obtained the phase diagram
in agreement with experiments.\cite{Shiina97,Shiba99}

Instead of pursuing this direction, we simply assume the ordering pattern, 
and calculate the RXS spectra.
The assumed ordering pattern selects a particular energy profile
according to Eq.~(\ref{eq.amplitude.E2.final}).
Note that the quartet $\Gamma_8$ consists of 16 degrees of freedom,
which are exhausted by three components of dipole,
five components of quadrupole and seven components of octupole operators
as well as an identical operator. Thereby the hexadecapole operators 
$H_4^0, H_4^2,H_4^4, H_x^{\beta}, H_y^{\beta}$
and $H_z^{\beta}$ are equivalent to
identical operator, $O_{x^2-y^2}, O_{3z^2-r^2}, O_{yz},
O_{zx}$ and $O_{xy}$, respectively, while
$H_{x,y,z}^{\alpha}=0$. Therefore, as long as a contribution from 
$\alpha_{E2}^{(2)}(\omega)$ exists, that from 
$\alpha_{E2}^{(4)}(\omega)$ automatically exists.

The order parameter in phase II is believed to be the $O_{xy}$-type.
Operator $O_{xy}$ has two degenerate eigenstates of eigenvalue $-1$ and 
two degenerate eigenstates of eigenvalue $+1$, that is,
\begin{eqnarray}
O_{xy} &=& \left( \begin{array}{cccc}
-1 &  0 & 0 & 0\\
 0 & -1 & 0 & 0 \\
 0 &  0 & 1 & 0 \\
 0 &  0 & 0 & 1 \\
\end{array} \right),
\end{eqnarray}
within the bases of eigenfunctions.
The AFQ phase may be constructed by assigning two degenerate eigenstates 
with eigenvalue $-1$ to one sublattice and those with 
eigenvalue $+1$ to the other sublattice. 
The degeneracy of the Kramers doublet would be lifted in the AFM phase
with further reducing temperatures.
Within the same bases in order, typical dipole and octupole operators
are represented by
\begin{eqnarray}
J_{z} &=& \left( \begin{array}{cccc}
-\frac{7}{6} & 0 & 0 &-\frac{2}{3} \\
 0 & \frac{7}{6} &-\frac{2}{3} & 0 \\
 0 &-\frac{2}{3} & \frac{7}{6} & 0 \\
-\frac{2}{3} & 0 & 0 &-\frac{7}{6} \\
\end{array} \right), \\
T_{z}^{\beta} &=& \left( \begin{array}{cccc}
 0 & 0 & 0 & {\textrm i} 3\sqrt{5} \\
 0 & 0 &-{\textrm i} 3\sqrt{5} & 0 \\
 0 & {\textrm i} 3\sqrt{5} & 0 & 0 \\
-{\textrm i} 3\sqrt{5} & 0 & 0 & 0 \\
\end{array} \right).
\end{eqnarray}
These forms indicate that the $O_{xy}$ order could accompany
neither the $J_z$ order nor the $T_{z}^{\beta}$ order.
Therefore, the $O_{xy}$ order selects the energy profiles
$\alpha^{(2)}_{E2}(\omega)$ and $\alpha^{(4)}_{E2}(\omega)$ 
according to Eq.~(\ref{eq.amplitude.E2.final}).

In the actual calculation of energy profile, 
we take into account full Coulomb interactions between 
$2p$ and $4f$ electrons, between $2p$ electrons,
and between $4f$ electrons in the configuration $(2p)^5(4f)^2$.
The spin-orbit interaction (SOI) of $2p$ and $4f$ electrons are
considered too.
The Slater integrals necessary for the Coulomb interactions
and the SOI parameters are evaluated within 
the Hartree-Fock approximation.\cite{Cowan81,com4}

\begin{figure}
\includegraphics[width=8.0cm]{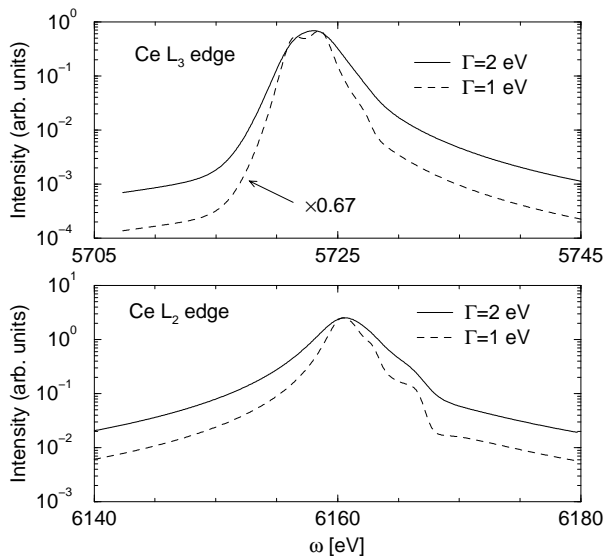}
\caption{\label{fig.CeB6.spec}
The RXS spectra at the Ce $L_3$ (top) and $L_2$ (bottom) absorption
edges from the AFQ phase (phase II).
$\Gamma=2.0$ eV and $1.0$ eV, ${\textbf G}=\left(\frac{3}{2} \frac{3}{2}
\frac{3}{2} \right)$, and $\psi=0$.  Only the spectra in the
$\sigma$-$\sigma'$ channel are displayed.
}
\end{figure}

Figure \ref{fig.CeB6.spec} shows the RXS spectra as a function of photon
energy, calculated with the core-hole lifetime 
width $\Gamma=2.0$ eV and 1.0 eV.
The energy of the Ce $2p$-core level is chosen 
such that the peak of the RXS spectra at the Ce L$_3$ edge 
coincides with the experiment for CeB$_{6}$.
We find that the absolute value of $\alpha^{(4)}_{E2}(\omega)$ is much smaller 
than that of $\alpha^{(2)}_{E2}(\omega)$. However, the smallness is compensated
by a large value of $\langle \psi_0|z_{\lambda}^{(4)}|\psi_0\rangle$,
and thereby both terms contribute to the intensity.
The calculated spectra show asymmetry and some structures,
which depend on the $\Gamma$ value.
It may be appropriate to use $\Gamma=2.0$ eV.\cite{Chen81}

\begin{figure}
\includegraphics[width=7.50cm]{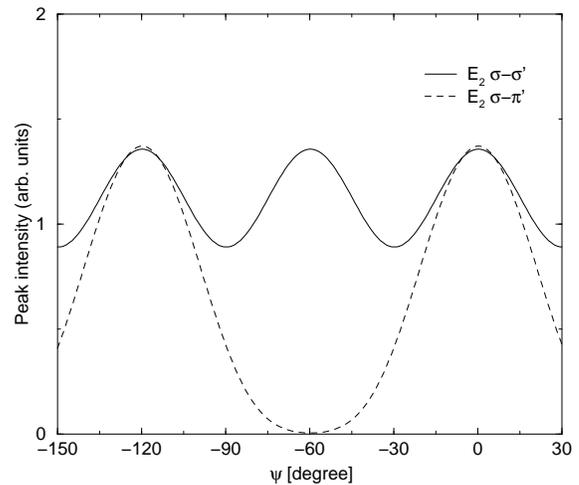}
\caption{\label{fig.CeB6.azim}
The peak intensities of RXS as functions of azimuthal angle
at the Ce $L_{3}$ edge from the AFQ pahse (phase II).
$\Gamma=2.0$ eV. Three domains are assumed to have equal volumes.
The solid and broken lines represent the $\sigma$-$\sigma'$
and $\sigma$-$\pi'$ channels, respectively.
}
\end{figure}

When the $O_{xy}$ order is realized, the $O_{yz}$ and $O_{zx}$ orders are 
also possible to be realized. In actual crystals, three orders may constitute 
domains, whose structure affects the azimuthal angle
dependence of the RXS spectra.  
Figure \ref{fig.CeB6.azim} shows the peak intensity 
as a function of $\psi$ for the scattering vector 
${\bf G}=\left(\frac{3}{2} \frac{3}{2} \frac{3}{2}\right)$.
The origin of $\psi$ is defined such that the scattering plane 
includes the $a$-axis.
It depends strongly on domains.
An incoherent addition over the contributions from three domains 
is performed.
In the $\sigma$-$\sigma'$ channel, the term of $\nu=2$ 
in Eq.~(\ref{eq.amplitude.E2.final}) is independent of $\psi$
so that the sixfold symmetry comes from the term of $\nu=4$.
On the other hand, the threefold symmetry in the $\sigma$-$\pi'$ channel
arises from both the term of $\nu=2$ and that of $\nu=4$.

\subsection{Phase IV in Ce$_{1-x}$La$_{x}$B$_{6}$}
 
The La diluted material Ce$_{1-x}$La$_{x}$B$_{6}$ with $x \simeq 0.3 \sim 0.5$
has an additional phase IV whose order parameter is not well 
understood yet.\cite{Tayama97}
Although a large discontinuity in the specific heat curve
suggests the existence of the long range order,\cite{Furuno85}
no neutron scattering experiment has found an evidence of long range
magnetic order.\cite{Takigawa02,Fischer05}
It is suggested\cite{Kubo03,Kubo04} that the AFO order
characterizes phase IV, which is supported by the observation
of the trigonal distortion.\cite{Akatsu03}
Recently, Mannix \textit{et al}. measured the RXS spectra at the $L_2$ edge 
in the $E$2 transition in Ce$_{0.7}$La$_{0.3}$B$_{6}$,
claiming that the signal arises from the AFO order.\cite{Mannix05}
The analysis of the azimuthal angle dependence by Kusunose and Kuramoto
supports the AFO order in phase IV.\cite{Kusunose05}
However, there exists at least one prominent discrepancy between
the experiment and the theory about the 
azimuthal angle dependence which we shall address later.

Keeping two possibilities, the quadrupole and octupole orders, 
for phase IV, we analyze the spectra on the basis of 
Eq.~({\ref{eq.amplitude.E2.final}).
Since the trigonal distortion is observed along the body-diagonal direction,
we assume that the order parameter is of $T^{\beta}_{111}$ type
($ T_{111}^{\beta} \equiv
 (T_{x}^{\beta}+T_{y}^{\beta}+T_{z}^{\beta})/\sqrt{3}$)
or $O_{111}$ type ($ O_{111}\equiv (O_{xy}+O_{yz}+O_{zx})/\sqrt{3}$).
The $T^{\alpha}$ type can be ruled out because this type carries
a substantial antiferromagnetic moment, which is against the experimental
finding.
Since $[T^{\beta}_{111},O_{111}]=0$, both operators are simultaneously
diagonalized. Within the bases of eigenfunctions, they are
represented as
\begin{eqnarray}
T_{111}^{\beta} &=& \left( \begin{array}{cccc}
 -3\sqrt{10} & 0 & 0 & 0 \\
  0 & 3\sqrt{10} & 0 & 0 \\
   0 & 0 & 0 & 0 \\
    0 & 0 & 0 & 0 \\
 \end{array} \right), \\
O_{111} &=& \left( \begin{array}{cccc}
    -1 & 0 & 0 & 0 \\
     0 &-1 & 0 & 0 \\
      0 & 0 & 1 & 0 \\
       0 & 0 & 0 & 1 \\
       \end{array} \right).
\label{eq.o111}
\end{eqnarray}
Within the same bases, the dipole operator $J_{111}$ 
($\equiv (J_x+J_y+J_z)/\sqrt{3}$)
is represented as
\begin{equation}
    J_{111} = \left( \begin{array}{cccc}
               0 & z_1 & 0 & 0 \\
      z_1^{\ast} & 0 & 0 & 0 \\
               0 & 0 &-\frac{7}{6} & 0 \\
               0 & 0 & 0 & \frac{7}{6} \\
	\end{array} \right), 
\label{eq.j111}
\end{equation}
where $z_1=\frac{\sqrt{3}}{18}(1 +{\textrm i}11\sqrt{2})$.

\subsubsection{AFO order}

The AFO order may be constructed by assigning the eigenstate of 
the $T_{111}^{\beta}$ with eigenvalue $-3\sqrt{10}$ to one sublattice and that 
with $3\sqrt{10}$ to the other sublattice. 
Then, the order parameter vector $(\langle  T_{x}^{\beta} \rangle,
\langle  T_{y}^{\beta} \rangle, \langle  T_{z}^{\beta} \rangle)$
is pointing to the $(111)$ direction.
Equation (\ref{eq.j111}) indicates that the AFO order could carry no magnetic 
moment, which is consistent with the experiment.
Equation (\ref{eq.o111}) indicates that the AFO order accompanies 
the ferroquadrupole order, not the AFQ order.
Therefore, according to Eq.~(\ref{eq.amplitude.E2.final}), the RXS energy 
dependence is purely characterized by $|\alpha_{E2}^{(3)}(\omega)|^2$.

Figure \ref{fig.CeLaB6.spec} shows the calculated 
$|\alpha_{E2}^{(3)}(\omega)|^2$ as functions of the incident photon energy 
$\omega$ at the Ce $L_2$ and $L_3$ absorption edges 
with $\Gamma=2.0$ eV and $1.0$ eV, 
in comparison with the experiment of Mannix \textit{et al}
(the non-resonant contribution is subtracted from the data).
\cite{Mannix05}
In the calculation, we use the same Slater integrals and the SOI parameters 
as in phase II. The spectral shapes depend strongly on the absorption edge 
they are observed.
In particular, the tail part of the spectra at the $L_3$ edge 
is drastically different from that at the $L_2$ edge. 
This fact might be helpful to identify the character of the
ordering pattern if the spectrum at the $L_3$ edge is experimentally
available.
Since the peak intensity at the $L_3$ edge is about 20 \%
of that at the $L_{2}$ edge, it can be said that
experimental observation has a legitimate chance at the former edge.
The $L_2$ spectral shape reproduces well the experimental one
showing broad single peak structure with a hump in the high energy region.

The energy profile $|\alpha_{E2}^{(3)}(\omega)|^2$ looks similar 
to the spectral shape in phase II (Fig. \ref{fig.CeB6.spec}) 
for $\Gamma=2$ eV. One difference is a dip found at the $L_3$ edge 
in $|\alpha_{E2}^{(3)}(\omega)|^2$, which is absent 
in Fig. \ref{fig.CeB6.spec}.
If the  $\Gamma$ is as small as $1$ eV,
the differences are emphasized around the tail part of the high energy 
region, because multiplet structures of the intermediate state are emphasized.
\cite{Nagao05}
Note that, although $|\alpha_{E2}^{(3)}(\omega)|^2$
is about two order of magnitude smaller than 
$|\alpha_{E2}^{(2)}(\omega)|^2$, the smallness is compensated by the large
factor of $|\langle T_{x,y,z}^{\beta} \rangle|^2 \simeq 90$,
resulting in the same order of magnitude of the spectral intensity
as in phase II (Fig. \ref{fig.CeB6.spec}).

\begin{figure}
\includegraphics[width=8.0cm]{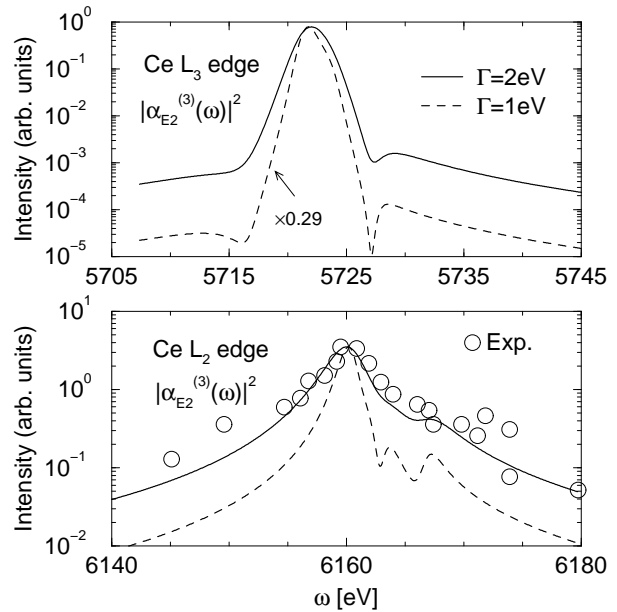}
\caption{\label{fig.CeLaB6.spec}
Energy profile  $|\alpha_{E2}^{(3)}(\omega)|^2$ 
at the Ce $L_3$ (top) and $L_2$ (bottom) absorption edges 
at ${\textbf G}=\left(\frac{3}{2} \frac{3}{2} \frac{3}{2} \right)$.  
Curves with $\Gamma=1.0$ eV (broken lines) are multiplied by factors
0.29 at the $L_3$ edge and ... at the $L_2$ edge
to have the same peak intensities as those with $\Gamma=2.0$ eV
(solid lines). Open circles are experimental data 
in Ce$_{0.7}$La$_{0.3}$B$_{6}$, in which non-resonant contributions are 
subtracted as explained by the authors.
\cite{Mannix05} 
}
\end{figure}

If the octupole order parameter vector $(\langle  T_{x}^{\beta} \rangle,
\langle  T_{y}^{\beta} \rangle, \langle  T_{z}^{\beta} \rangle)$ can point
to the $(111)$ direction, it is also possible to
point to the $(1\overline{11})$, $(\overline{1}1\overline{1})$, 
and $(\overline{11}1)$ directions. 
These four orders usually constitute domains.
The azimuthal angle dependence is different for different domains,
as shown in Figs.~\ref{fig.CeLaB6.azim}(a) and (b).
If you collect the contributions from domains with equal weight, 
the maximum intensity in the $\sigma$-$\pi'$ channel become nearly equal 
to that in the $\sigma$-$\sigma'$ channel.
The experimental data show that the maximum intensity 
in the $\sigma$-$\pi'$ channel is about the half of that 
in the $\sigma$-$\sigma'$ channel, as shown in Fig.~\ref{fig.CeLaB6.azim}(c).
This may be attributed to the slightly
different setup for different polarizations and/or to the extrinsic
background from the non-resonant contribution,
as discussed by Kusunose and Kuramoto.\cite{Kusunose05}
They reduced the intensity in the $\sigma$-$\pi'$ channel
by simply multiplying a factor $0.6$.
Another possibility is that domain volumes are different among four domains.
Collecting up the contributions with ratio $3:1:1:1$ from the
$(111)$, $(1\overline{11})$, $(\overline{1}1\overline{1})$, 
and $(\overline{11}1)$ domains, we have the result
similar to that simply multiplying a factor $0.6$ to the intensity 
in the $\sigma$-$\pi'$ channel, as shown in Fig.~\ref{fig.CeLaB6.azim}(c).
Thus, the sixfold and threefold symmetries in the 
$\sigma$-$\sigma'$ and $\sigma$-$\pi'$ channels
are well reproduced in comparison with the experiment. 

\begin{figure}
\includegraphics[width=8.0cm]{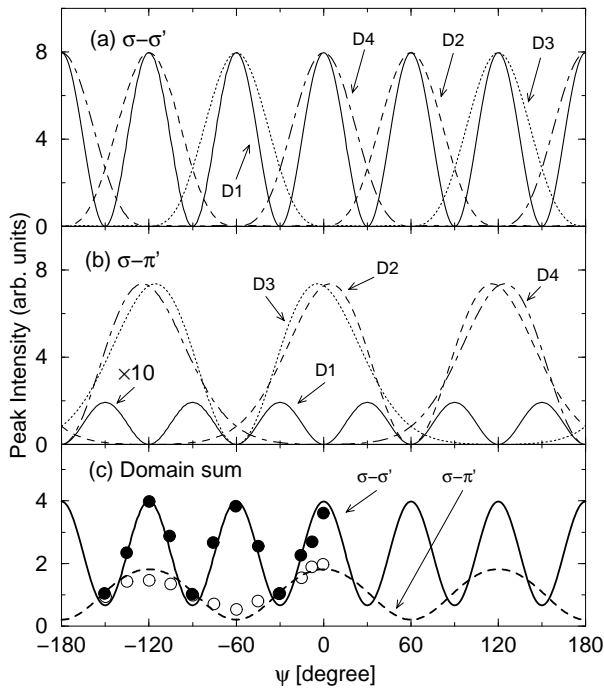}
\caption{\label{fig.CeLaB6.azim}
Peak intensities of the RXS spectra at the Ce $L_{2}$ edges 
in the AFO phase as functions of azimuthal angle.
Panels (a) and (b) display the peak intensities in the
$\sigma$-$\sigma'$ and the $\sigma$-$\pi'$ channels,
respectively, where
the solid (D1), broken (D2), dotted (D3), and broken-dotted (D4) lines
represent the peak intensity of the domains $(111)$, $(1\overline{11})$, 
$(\overline{1}1\overline{1})$, and $(\overline{11}1)$, respectively.
Panel (c) shows the intensities collecting
the contributions from the domains $(111)$, $(1\overline{11})$, 
$(\overline{1}1\overline{1})$, and $(\overline{11}1)$ with ratio 3:1:1:1. 
The solid and broken lines represent the intensities in the $\sigma$-$\sigma'$
and $\sigma$-$\pi'$ channels, respectively.
Filled and open circles are the experimental data for 
Ce$_{0.7}$La$_{0.3}$B$_{6}$.
\cite{Mannix05}
}
\end{figure}

\subsubsection{AFQ order}

The AFQ order may be constructed by assigning the eigenstates of $O_{111}$
with eigenvalue $-1$ to one sublattice and those with $+1$ to the other 
sublattice.
The AFQ order accompanies no AFO order.
The difference from phase II is that the order parameter 
$(\langle O_{yz}\rangle,\langle O_{zx}\rangle, \langle O_{xy}\rangle)$
is pointing to the $(111)$ direction.
Therefore, the spectral shape as a function of energy is nearly the same as
in phase II.
The azimuthal angle dependence depends strongly on domains,
which is shown in Figs. ~\ref{fig.CeLaB6.azim.2}(a) and (b).
The sixfold symmetry in the $\sigma$-$\sigma'$ channel
mainly comes not from the $(111)$ domain but from the other domains,
since the contribution from the former domain is constant
with varying the azimuthal angle. 
Collecting the contributions from four domains with
equal weight, and reducing the intensity in the $\sigma$-$\pi'$ channel
by multiplying a factor $0.6$ in the same way as Kusunose and Kuramoto
adopted,\cite{Kusunose05}
we obtain the result in agreement with the experiment
at least in a symmetrical point of view.
Although the amplitude of the oscillation in the $\sigma$-$\sigma'$
channel is too small compared with the experimental one,
the situation may be changed if the subtraction of the
non-resonant contribution in the $\sigma$-$\sigma'$ channel
and/or that of the enigmatic $E$1 contribution in the
$\sigma$-$\pi'$ channel are/is reevaluated.
Actually, the discrepancy about the relative intensity between two
channel may be attributed to the consequence of these subtraction process.

We now turn to our attention to the energy dependence of the spectra.
Owing to our formula Eq. (\ref{eq.amplitude.E2.final}),
the spectral shapes from the AFQ phase with $O_{111}$ type
in phase IV are the same as those with $O_{xy}$ type in phase II 
(Fig.~\ref{fig.CeB6.spec}). Therefore, the energy dependence 
at the $L_2$ edge is similar to that obtained from the AFO phase.
On the other hand, the energy dependence at the $L_3$edge from the AFQ phase
is different from that obtained from the AFO phase,
which may help the identification of the ordering pattern
realized in this material.

\begin{figure}
\includegraphics[width=8.0cm]{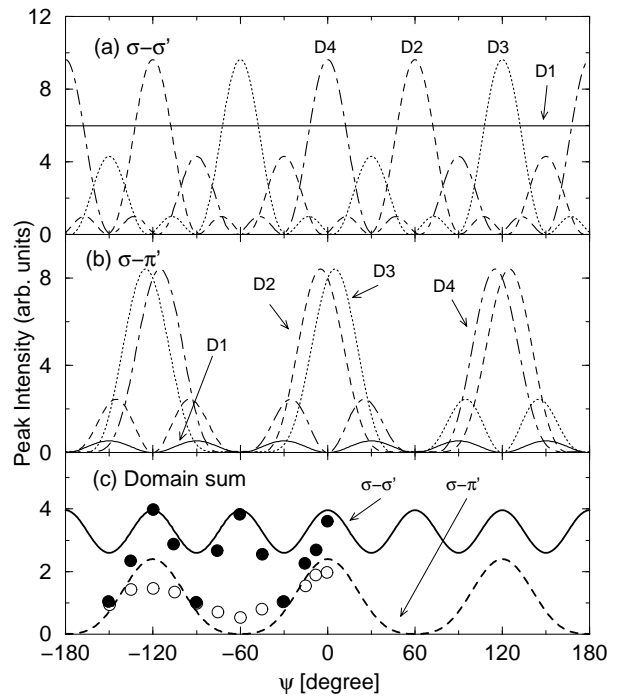}
\caption{\label{fig.CeLaB6.azim.2}
Peak intensities at the Ce $L_{2}$ edges in the AFQ phase,
as functions of azimuthal angle.
Panels (a) and (b) display the peak intensities in the
$\sigma$-$\sigma'$ and the $\sigma$-$\pi'$ channels,
respectively, where
the solid (D1), broken (D2), dotted (D3), and broken-dotted (D4) lines
represent the peak intensity of the domains $(111)$, 
$(1\overline{11})$, $(\overline{1}1\overline{1})$, and 
$(\overline{11}1)$, respectively.
Panel (c) shows the result collecting
the contributions from the domains with equal weight.
The solid and broken lines represent the $\sigma$-$\sigma'$
and $\sigma$-$\pi'$ channels, respectively.
The curve in the latter channel is multiplied by a factor 0.6.
Filled and open circles are the experimental data
for Ce$_{0.7}$La$_{0.3}$B$_{6}$.\cite{Mannix05}
}
\end{figure}

\section{\label{sect.4} Concluding remarks}

We have derived a general formula
of the RXS amplitude in the $E$2 transition.
The derivation is based on the assumption that
the Hamiltonian describing the intermediate state of
the scattering process preserves the spherical symmetry.
The obtained formula is applicable to many $f$ electron systems
where a localized scheme gives a good description.
Although similar formulae have already 
been obtained,\cite{Hannon88,Luo93,Carra94,Hill96,Lovesey96}
the present formula has two prominent advantages.
One is that it is able to calculate the energy profile of the RXS spectra,
because our treatment is free from the fast collision 
approximation adopted in the previous works.
The other is that it is conveniently applicable to the 
systems possessing multipole order parameters.
 
We have demonstrated the usefulness of the derived formula by calculating
the $E$2 RXS spectra in Ce$_{1-x}$La$_{x}$B$_{6}$.
Phase II is believed to be an AFQ order of $O_{xy}$ type,
and our formula dictates that the energy dependence is given by a combination 
of $\alpha_{E2}^{(2)}(\omega)$ and $\alpha_{E2}^{(4)}(\omega)$. 
We have obtained the RXS intensity in the same order of intensity as
obtained by assuming the AFO order. This suggests that the $E$2 signal 
is detectable from phase II, 
although only the $E$1 signal has been reported
in phase II of CeB$_6$.\cite{Nakao01,Yakhou01}
Subsequently, we have calculated the RXS spectra by assuming 
the $T_{111}^{\beta}$-type AFO order, in order to clarify the order parameter 
of phase IV.  The energy dependence $|\alpha_{E2}^{(3)}(\omega)|^2$
has been obtained at the $L_{2}$ edge in agreement with the experiment 
in Ce$_{0.7}$La$_{0.3}$B$_{6}$.\cite{Mannix05}
Unfortunately this is not used
to discriminate between the AFO and AFQ orders, because  
the spectral shapes are nearly the same in the two ordering phases.
On the other hand, the spectral shape at the $L_3$ edge has been found
slightly different from the $L_2$ edge,
which might help the identification of the ordering pattern.  
For the azimuthal angle dependence, we have reproduced
the sixfold and threefold symmetries by assuming the AFO order,
in agreement with the previous theoretical analysis and the 
experiment.\cite{Mannix05,Kusunose05}
The intensity in the $\sigma$-$\pi'$ channel becomes nearly equal to 
that in the $\sigma$-$\sigma'$ channel with the equal volume for four domains,
while in the experiment the intensity in the former channel is found
nearly half of that in the latter. This discrepancy may be removed by
assuming uneven volumes among four domains. 
We have also analyzed the azimuthal angle dependence by assuming
the $O_{111}$-type AFQ order. It is found that the simultaneously induced 
hexadecapole order gives rise to the sixfold and threefold symmetries.
Although the agreement with the experiment is quantitatively not good,
it may be difficult to rule out the AFQ order from phase IV
on the basis of the azimuthal angle dependence alone.
Since it depends strongly on the domain distribution,
experiments controlling the domain distribution, if possible,
might be useful to clarify the situation.

\begin{acknowledgments}
We thank M. Takahashi and M. Yokoyama for valuable discussions.
This work was partially supported by a Grant-in-Aid for Scientific Research 
from the Ministry of Education, Science, Sports and Culture, Japan.
\end{acknowledgments}

\appendix

\section{\label{app.A} Definitions of some quantities used 
in Sec. \ref{sect.2}}

Let us define irreducible tensor operator of rank $\nu$
with the spherical basis. The $n$-th component
($-\nu \leq n \leq \nu$) $T_{n}^{(\nu)}$ is defined recurrently as
\begin{eqnarray}
T_{\nu}^{(\nu)} &=& (-)^{\nu} \sqrt{\frac{(2\nu-1)!!}{(2\nu)!!}} J_+^{\nu}, \\
\ [J_-,T_{n}^{(\nu)}] &=& \sqrt{(\nu+n)(\nu-n+1)} T_{n-1}^{(\nu)}.
\end{eqnarray}

\begin{table}
\caption{\label{table.9}
Irreducible tensor operator $T_{n}^{(\nu)}$ with the spherical basis.
}
\begin{ruledtabular}
\begin{tabular}{crl}
rank & & \\
 $\nu$ & $n$ & $T_n^{(\nu)}$ \\
\hline
1        & $\pm$ 1  & $\mp \frac{1}{\sqrt{2}}J_{\pm}$ \\
         &  0  & $J_z$ \\
\hline
2        & $\pm$ 2  & $\frac{1}{2}\sqrt{\frac{3}{2}} J_{\pm}^2$ \\
         & $\pm$ 1  & $\mp \frac{1}{2}\sqrt{\frac{3}{2}} 
                       J_{\pm} (2 J_z \pm 1)$ \\
         &     0  & $\frac{1}{2} [3 J_z^2 -J(J+1) ]$ \\
\hline
3        & $\pm$ 3  & $\mp \frac{\sqrt{5}}{4} J_{\pm}^3$ \\
         & $\pm$ 2  & $\frac{\sqrt{15}}{2\sqrt{2}} J_{\pm}^2 (J_z \pm 1)$ \\
         & $\pm$ 1  & $\mp \frac{1}{4\sqrt{3}} J_{\pm} 
         [ 15 J_z^2 \pm 15 J_z - 3J(J+1) + 6]$ \\
         &     0  & $\frac{1}{2} [ 5 J_z^3 - 3J(J+1)J_z + J_z]$ \\
\hline
4        & $\pm$ 4  & $\frac{\sqrt{35}}{8\sqrt{2}}  J_{\pm}^4$ \\
         & $\pm$ 3  & $\mp \frac{\sqrt{35}}{8} J_{\pm}^3(2J_z \pm 3)$ \\
         & $\pm$ 2  & $\frac{\sqrt{5}}{4\sqrt{2}} J_{\pm}^2[7J_z^2 \pm 14 J_z
                       -J(J+1) +9]$ \\
         & $\pm$ 1  & $\mp \frac{\sqrt{5}}{8} J_{\pm}
                       [14J_z^3 \pm 21 J_z^2+19 J_z$ \\
         &          & $-6J(J+1)J_z \mp 3J(J+1) \pm 6]$ \\
         &       0  & $\frac{1}{8} [35J_z^4 - 30J(J+1) J_z^2+25 J_z^2$ \\
         &          & $+3J^2(J+1)^2 -6J(J+1)]$ \\
\end{tabular}
\end{ruledtabular}
\end{table}
Expressions for $T_{n}^{(\nu)}$'s are listed in Table \ref{table.9} 
up to rank four.
We can find $(2\nu +1) \times (2 \nu+1)$ unitary matrix
which connects
the tensor operator with the spherical component $T_{n}^{(\nu)}$
and that with the Cartesian component $z_{\lambda}^{(\nu)}$
which satisfies
\begin{equation}
z_{\lambda}^{(\nu)} = \sum_{n=\nu}^{\nu}
U_{\lambda n}^{(\nu)} T_{n}^{(\nu)},
\end{equation}
and inversely,
\begin{equation}
T_{n}^{(\nu)} = \sum_{\lambda=1}^{2\nu+1}
[U^{(\nu) \dagger}]_{n \lambda} z_{\lambda}^{(\nu)}.
\end{equation}
Explicit form of $U^{(\nu)}$ is summarized in Table \ref{table.10}.

\squeezetable
\begin{table}
\caption{\label{table.10}
Unitary matrix which connects the tensor operator with the Cartesian basis
and that with the spherical basis.
}
\begin{ruledtabular}
\begin{tabular}{cc}
$U^{(0)}$ & 1 \\
\hline
$U^{(1)}$ & $\left( \begin{array}{ccc}
  -\frac{1}{\sqrt{2}} & 0 & \frac{1}{\sqrt{2}} \\
   \frac{\rm i}{\sqrt{2}} & 0 & \frac{\rm i}{\sqrt{2}} \\
   0 & 1 & 0 \\
\end{array} \right)$ \\
\hline
$U^{(2)}$ & $\left( \begin{array}{ccccc}
   \frac{1}{\sqrt{2}} & 0 & 0 & 0 & \frac{1}{\sqrt{2}} \\
   0 & 0 & 1 & 0 & 0  \\
   0 & \frac{\rm i}{\sqrt{2}} & 0 & \frac{\rm i}{\sqrt{2}} & 0 \\
   0 &-\frac{1}{\sqrt{2}} & 0 & \frac{1}{\sqrt{2}} & 0 \\
  -\frac{\rm i}{\sqrt{2}} & 0 & 0 & 0 & \frac{\rm i}{\sqrt{2}} \\
\end{array} \right)$ \\
\hline
$U^{(3)}$ & $\left( \begin{array}{ccccccc}
   0 &-\frac{\rm i}{\sqrt{2}} & 0 & 0 & 0 & \frac{\rm i}{\sqrt{2}} & 0 \\
  -\frac{\sqrt{5}}{4} & 0 & \frac{\sqrt{3}}{4} & 0 &-\frac{\sqrt{3}}{4} &
  0 & \frac{\sqrt{5}}{4} \\
  -\frac{\sqrt{5}}{4}{\rm i} & 0 &-\frac{\sqrt{3}}{4}{\rm i} & 0 
  &-\frac{\sqrt{3}}{4}{\rm i} & 0 &-\frac{\sqrt{5}}{4}{\rm i} \\
   0 & 0 & 0 & 1 & 0 & 0 & 0  \\
   \frac{\sqrt{3}}{4} & 0 & \frac{\sqrt{5}}{4} & 0 &-\frac{\sqrt{5}}{4} &
  0 &-\frac{\sqrt{3}}{4} \\
  -\frac{\sqrt{3}}{4}{\rm i} & 0 & \frac{\sqrt{5}}{4}{\rm i} & 0 
  & \frac{\sqrt{5}}{4}{\rm i} & 0 &-\frac{\sqrt{3}}{4}{\rm i} \\
   0 & \frac{1}{\sqrt{2}} & 0 & 0 & 0 & \frac{1}{\sqrt{2}} & 0 \\
\end{array} \right)$ \\
\hline
$U^{(4)}$ & $\left( \begin{array}{ccccccccc}
     \frac{\sqrt{30}}{12} & 0 & 0 & 0 & \frac{\sqrt{21}}{6} &
     0 & 0 & 0 & \frac{\sqrt{30}}{12} \\
     0 & 0 &-\frac{1}{\sqrt{2}} & 0 & 0 & 0 &-\frac{1}{\sqrt{2}} & 0 & 0 \\
    -\frac{\sqrt{42}}{12} & 0 & 0 & 0 & \frac{\sqrt{15}}{6} &
     0 & 0 & 0 &-\frac{\sqrt{42}}{12} \\
     0 &-\frac{\rm i}{4} & 0 &-\frac{\sqrt{7}}{4}{\rm i} & 0 &
    -\frac{\sqrt{7}}{4}{\rm i} & 0 &-\frac{\rm i}{4} & 0 \\
     0 & \frac{1}{4} & 0 &-\frac{\sqrt{7}}{4} & 0 &
     \frac{\sqrt{7}}{4} & 0 &-\frac{1}{4} & 0 \\
    -\frac{\rm i}{\sqrt{2}} & 0 & 0 & 0 & 0 & 0 & 
     0 & 0 & \frac{\rm i}{\sqrt{2}}  \\
     0 & \frac{\sqrt{7}}{4}{\rm i} & 0 &-\frac{\rm i}{4} & 0 &
    -\frac{\rm i}{4} & 0 & \frac{\sqrt{7}}{4}{\rm i} & 0 \\
     0 & \frac{\sqrt{7}}{4} & 0 & \frac{1}{4} & 0 &
    -\frac{1}{4} & 0 &-\frac{\sqrt{7}}{4} & 0 \\
     0 & 0 &-\frac{\rm i}{\sqrt{2}} & 0 & 0 & 
     0 & \frac{\rm i}{\sqrt{2}} & 0 & 0 \\
\end{array} \right)$ \\
\end{tabular}
\end{ruledtabular}
\end{table}

Finally, we show the explicit forms of 
the functions $\alpha_{E2}^{(\nu)} (\omega)$, which give
the energy profiles coupled to the
expectation value of the rank-$\nu$ multipole operator.
\begin{eqnarray}
\alpha_{E2}^{(4)} (\omega) 
  &=& 8 \sqrt{\frac{2}{35}} \sum_{J'=J-2}^{J+2} F_{J'}(\omega),
\\
\alpha_{E2}^{(3)}(\omega) 
&=&
4\sqrt{\frac{2}{5}} \left[
    -(2J-3) F_{J-2} - (J-2) F_{J-1} \right. \nonumber \\
& & \left. + 2 F_{J} + (J+3) F_{J+1} 
 + (2J+5) F_{J+2} \right],
\\
\alpha_{E2}^{(2)}(\omega)
&=& 2 \sqrt{\frac{2}{7}} \left[
    4(2J-3)(J-1) F_{J-2} \right. \nonumber \\
& & +(J-5)(J-1) F_{J-1} \nonumber \\
& &  -\frac{1}{3}(2J-3)(2J+5) F_{J} 
+(J+2)(J+6) F_{J+1} \nonumber \\
& & \left. 
+4(2J+5)(J+2) F_{J+2} \right],
\\
\alpha_{E2}^{(1)}(\omega) 
&\equiv&-\sqrt{\frac{2}{5}}[4(J-1)(2J-1)(2J-3)F_{J-2} \nonumber \\
& & -(J-1)(2J-1)(J+3) F_{J-1} \nonumber \\
& & + (2J-1)(2J+3) F_{J} \nonumber \\
& & + (J+2)(J-2)(2J+3) F_{J+1} \nonumber \\
& & - 4(J+2)(2J+3)(2J+5) F_{J+2} ],
\\
\alpha_{E2}^{(0)}(\omega) 
&\equiv& \frac{2}{3\sqrt{5}} \left[ 6J(J-1)(2J-1)(2J-3) F_{J-2} \right.
\nonumber \\ 
& & - 3J(J-1)(J+1)(2J-1) F_{J-1} \nonumber \\ 
& & + J(J+1)(2J-1)(2J+3) F_{J} \nonumber \\ 
& & - 3J(J+1)(J+2)(2J+3) F_{J+1} \nonumber \\ 
& & \left. + 6(J+1)(J+2)(2J+3)(2J+5) 
   F_{J+2} \right]. \nonumber \\
\end{eqnarray}
The energy dependence is contained in the functions $F_{J'}(\omega)$
as
\begin{eqnarray}
F_{J'}(\omega)&=& _{4}{\textrm C}_{J-J'+2} \sqrt{(2J+1)(2J'+1)}
\frac{(J+J'-2)!}{(J+J'+3)!} \nonumber \\
&\times& |(J||V_2||J')|^2 \sum_{i=1}^{N_{J'}} E_i(\omega,J'),
\end{eqnarray}
where $_{n}{\textrm C}_{m}=\frac{n!}{m!(n-m)!}$ represents combination.

\section{\label{app.B}Geometrical factors}

The geometrical factors $P_{\mu}^{(\nu)}$ for $\nu=3$ and  $4$  
in Eq. (\ref{eq.amplitude.E2.final}) have rather complicated forms.
For $\nu=3$, they are summarized as follows:
\begin{eqnarray}
P_{1}^{(3)} &\equiv& 
{\textit i} \frac{1}{3\sqrt{2}}
\left[
      [\hat{\textbf k}' \times \hat{\textbf k}] \cdot 
   {\textbf q}(\mbox{\boldmath{$\epsilon$}}',\mbox{\boldmath{$\epsilon$}})
+ [\mbox{\boldmath{$\epsilon$}}' \times \mbox{\boldmath{$\epsilon$}}]
  \cdot {\textbf q}(\hat{\textbf k}',\hat{\textbf k})
\right. \nonumber \\
& & \left. +
 [\hat{\textbf k}' \times \mbox{\boldmath{$\epsilon$}}]
   \cdot {\textbf q}(\mbox{\boldmath{$\epsilon$}}',\hat{\textbf k})
+[\mbox{\boldmath{$\epsilon$}}' \times \hat{\textbf k}]
\cdot {\textbf q}(\hat{\textbf k}',\mbox{\boldmath{$\epsilon$}}) \right],
 \\
P_{\mu}^{(3)} 
&=& \frac{\textit i}{2} \sqrt{\frac{5}{2}} 
 \left[ [\hat{\textbf k}' \times \hat{\textbf k}]_{\mu} 
    \epsilon_{\mu}'\epsilon_{\mu}
+ [\mbox{\boldmath{$\epsilon$}}' \times \mbox{\boldmath{$\epsilon$}}]_{\mu}
 \hat{k}_{\mu}' \hat{k}_{\mu}
\right. \nonumber \\
& & \left. +
 [\hat{\textbf k}' \times \mbox{\boldmath{$\epsilon$}}]_{\mu}
   \epsilon_{\mu}' \hat{k}_{\mu}
+[\mbox{\boldmath{$\epsilon$}}' \times \hat{\textbf k}]_{\mu}
       \hat{k}_{\mu}'\epsilon_{\mu} \right]
 \nonumber \\
&+& \frac{\textit i}{2\sqrt{10}} P_{\mu}^{(1)} 
\quad  \quad \textrm{for}  \ \mu=2,3, \ \textrm{and} \   4,\\
P_{\mu}^{(3)}
&=& \frac{\textit i}{4}\sqrt{\frac{3}{2}}
(\hat{\textbf k}' \times \hat{\textbf k} )_{\mu}
\sum_{\mu',\mu''=5}^{7}\epsilon_{\mu \mu' \mu''}
 (\epsilon_{\mu'}' \epsilon_{\mu'} - \epsilon_{\mu''}'\epsilon_{\mu''} ) 
\nonumber \\
&+& \frac{\textit i}{4}\sqrt{\frac{3}{2}} 
(\mbox{\boldmath{$\epsilon$}}' \times \mbox{\boldmath{$\epsilon$}})_{\mu} 
\sum_{\mu',\mu''=5}^{7}\epsilon_{\mu \mu' \mu''}
 ( \hat{k}_{\mu'}' \hat{k}_{\mu'} - \hat{k}_{\mu''}' \hat{k}_{\mu''} ) \nonumber \\
&+& \frac{\textit i}{4}\sqrt{\frac{3}{2}} (\hat{\textbf k}' 
   \times \mbox{\boldmath{$\epsilon$}})_{\mu} 
\sum_{\mu',\mu''=5}^{7}\epsilon_{\mu \mu' \mu''}
 ( \epsilon_{\mu'}' \hat{k}_{\mu'} - \epsilon_{\mu''}' \hat{k}_{\mu''} ) 
\nonumber \\
&+& \frac{\textit i}{4}\sqrt{\frac{3}{2}}
(\mbox{\boldmath{$\epsilon$}}' \times \hat{\textbf k} )_{\mu}
\sum_{\mu',\mu''=5}^{7}\epsilon_{\mu \mu' \mu''}
 (\hat{k}_{\mu'}' \epsilon_{\mu'} - \hat{k}_{\mu''}'\epsilon_{\mu''} ) 
\nonumber \\
& &
\hspace*{3.0cm} \textrm{for} \ \mu =5,6, \ \textrm{and} \   7.
\end{eqnarray}
Note that $\mu=2, 3$, and $4$ work as $x, y$, and $z$, respectively.
Similarly, $\mu=5, 6$, and $7$ work as $x, y$, and $z$, respectively.
The Levi-Civita tensor density $\epsilon_{\mu \mu' \mu''}$
is introduced.

For $\nu=4$, the results are as follows.
\begin{eqnarray}
P_{1}^{(4)}
&=& \sqrt{\frac{2}{15}} 
[ 5(\hat{k}_{x}' \hat{k}_{x} \epsilon_{x}' \epsilon_{x}
+ \hat{k}_{y}' \hat{k}_{y} \epsilon_{y}' \epsilon_{y}
+ \hat{k}_{z}' \hat{k}_{z} \epsilon_{z}' \epsilon_{z} )
\nonumber \\
&-& P_{1}^{(0)} ], \\
P_{2}^{(4)}
&=&  \sqrt{14}
 ( \hat{k}_x' \hat{k}_x \epsilon_x' \epsilon_x
          - \hat{k}_y' \hat{k}_y \epsilon_y' \epsilon_y )  \nonumber \\
&-&2 \sqrt{\frac{2}{21}} \left[
      [\hat{\textbf k}' \cdot \hat{\textbf k}] 
q_{1}(\mbox{\boldmath{$\epsilon$}}',\mbox{\boldmath{$\epsilon$}})
+ [\mbox{\boldmath{$\epsilon$}}' \cdot \mbox{\boldmath{$\epsilon$}}]
q_{1}(\hat{\textbf k}',\hat{\textbf k}) \right.\nonumber \\
& & \left. +
 [\hat{\textbf k}' \cdot \mbox{\boldmath{$\epsilon$}}]
q_{1}(\mbox{\boldmath{$\epsilon$}}',\hat{\textbf k})
+[\mbox{\boldmath{$\epsilon$}}' \cdot \hat{\textbf k}]
q_{1}(\hat{\textbf k}',\mbox{\boldmath{$\epsilon$}}) \right],  \\
P_{3}^{(4)}
&=& -\sqrt{14} 
 [\hat{k}_{x}' \hat{k}_{x} \epsilon_{x}' \epsilon_{x}
 +\hat{k}_{y}' \hat{k}_{y} \epsilon_{y}' \epsilon_{y}
 -2 \hat{k}_{z}' \hat{k}_{z} \epsilon_{z}' \epsilon_{z} ]  \nonumber \\
&-& 2 \sqrt{\frac{2}{7}} 
\left[
      [\hat{\textbf k}' \cdot \hat{\textbf k}] 
q_{2}(\mbox{\boldmath{$\epsilon$}}',\mbox{\boldmath{$\epsilon$}})
+ [\mbox{\boldmath{$\epsilon$}}' \cdot \mbox{\boldmath{$\epsilon$}}]
q_{2}(\hat{\textbf k}',\hat{\textbf k}) \right.\nonumber \\
& & \left. +
 [\hat{\textbf k}' \cdot \mbox{\boldmath{$\epsilon$}}]
q_{2}(\mbox{\boldmath{$\epsilon$}}',\hat{\textbf k})
+[\mbox{\boldmath{$\epsilon$}}' \cdot \hat{\textbf k}]
q_{2}(\hat{\textbf k}',\mbox{\boldmath{$\epsilon$}}) \right], \\
P_{\mu}^{(4)}
&=& 
\frac{1}{2\sqrt{6}}
   q_{\mu+1}(\hat{\textbf k}',\hat{\textbf k})
   \sum_{\mu',\mu''=4}^{6} \epsilon_{\mu \mu' \mu''}
    (\epsilon_{\mu'}' \epsilon_{\mu'} - \epsilon_{\mu''}' \epsilon_{\mu''} )
\nonumber \\
&+& \frac{1}{2\sqrt{6}}
   q_{\mu+1}(\mbox{\boldmath{$\epsilon$}}',\mbox{\boldmath{$\epsilon$}})
   \sum_{\mu',\mu''=4}^{6} \epsilon_{\mu \mu' \mu''}
    (\hat{k}_{\mu'}'\hat{k}_{\mu'} - \hat{k}_{\mu''}' \hat{k}_{\mu''}) 
\nonumber \\
&+&
\frac{1}{2\sqrt{6}}
   q_{\mu+1}(\hat{\textbf k}',\mbox{\boldmath{$\epsilon$}})
   \sum_{\mu',\mu''=4}^{6} \epsilon_{\mu \mu' \mu''}
    (\epsilon_{\mu'}' \hat{k}_{\mu'} - \epsilon_{\mu''}' \hat{k}_{\mu''} )
\nonumber \\
&+& \frac{1}{2\sqrt{6}}
   q_{\mu+1}(\mbox{\boldmath{$\epsilon$}}',\hat{\textbf k})
   \sum_{\mu',\mu''=4}^{6} \epsilon_{\mu \mu' \mu''}
    (\hat{k}_{\mu'}' \epsilon_{\mu'} - \hat{k}_{\mu''}' \epsilon_{\mu''})
\nonumber \\
& & \hspace*{3.0cm} \textrm{for} \ \mu =4,5, \ \textrm{and} \   6,
\label{eq.B7} \\
P_{\mu}^{(4)}
&=& \frac{1}{\sqrt{42}} [ 7 \hat{k}_{\mu}' \hat{k}_{\mu} -3 (\hat{\textbf k}' \cdot \hat{\textbf k}) ]
  q_{\mu-4}(\mbox{\boldmath{$\epsilon$}}',\mbox{\boldmath{$\epsilon$}})
\nonumber \\
&+&\frac{1}{\sqrt{42}} [7 \epsilon_{\mu}' \epsilon_{\mu}
  -3(\mbox{\boldmath{$\epsilon$}}' \cdot \mbox{\boldmath{$\epsilon$}}) ]
  q_{\mu-4}(\hat{\textbf k}',\hat{\textbf k}) \nonumber \\
&+&\frac{1}{\sqrt{42}} [7 \epsilon_{\mu}' \hat{k}_{\mu}
             -3 (\mbox{\boldmath{$\epsilon$}}' \cdot \hat{\textbf k}) ]
  q_{\mu-4}(\hat{\textbf k}',\mbox{\boldmath{$\epsilon$}}) \nonumber \\
&+&\frac{1}{\sqrt{42}} [ 7 \hat{k}_{\mu}' \epsilon_{\mu}
                -3 (\hat{\textbf k}' \cdot \mbox{\boldmath{$\epsilon$}}) ]
  q_{\mu-4}(\mbox{\boldmath{$\epsilon$}}',\hat{\textbf k})
\nonumber \\
& & \hspace*{3.0cm} \textrm{for} \ \mu =7,8, \ \textrm{and} \   9,
\label{eq.B8} 
\end{eqnarray}
where indices 4, 5, and 6 in the summations
in Eq.~(\ref{eq.B7}) serve as $x, y$, and $z$, respectively.
Similarly, indices 7, 8, and 9 appeared in the brakets in Eq.~ (\ref{eq.B8})
serve as $x, y$, and $z$, respectively.

\bibliography{paper.E2.submit}

\begin{thebibliography}{40}
\bibitem[{com()}]{com4}
\bibinfo{note}{In solids, the magnitude of the Slater integrals are known to be
  reduced due to large screening effect. The isotropic and the anisotropic
  parts of the Slater integral are used by multiplying factors 0.25 and 0.8,
  respectively.}
\end{thebibliography}

\end{document}